\documentclass[preprint,twocolumn]{rmaa}

\title{An atlas of synthetic line profiles of planetary nebulae }

\author{C.~Morisset \altaffilmark{1}
   and G.~Stasi\'nska \altaffilmark{2}
}

\altaffiltext{1}{Instituto de Astronom{\'\i}a, Universidad Nacional 
Aut\'onoma de M\'exico, Apdo. Postal 70264, M\'ex. D.F., 04510 
M\'exico, Morisset@Astroscu.UNAM.mx}
\altaffiltext{2}{LUTH, Observatoire de Paris, CNRS, Universit\'e 
Paris Diderot; Place Jules Janssn 92190 Meudon, France }

\shortauthor{Morisset \& Stasi\'nska}
\shorttitle{Atlas of synthetic nebular line profiles}

\fulladdresses{
\item C. Morisset: Instituto de Astronom\'{\i}a,
   Universidad Nacional Aut\'onoma de M\'exico
   Apdo. postal 70--264; Ciudad Universitaria;
   M\'exico D.F. 04510; M\'exico.
\item G.~Stasi\'nska : LUTH, Observatoire de Paris, CNRS, Universit\'e Paris Diderot; Place Jules Janssen 92190 Meudon, France.\\
}

\newcommand{\mysp}{}\def\mysp/{}
\newcommand{\teff}{}\def\teff/{$\mathrm{T}_{\mathrm{eff}}$}
\newcommand{\kms}{}\def\kms/{km.s$^{-1}$}
\newcommand{\hbeta}{}\def\hbeta/{H$\beta$}
\newcommand{\hei}{}\def\hei/{\ion{He\mysp/}{I}\ 5876\AA}
\newcommand{\heii}{}\def\heii/{\ion{He\mysp/}{II}\ 4686\AA}
\newcommand{\nii}{}\def\nii/{[\ion{N\mysp/}{II}]\ 6584\AA}
\newcommand{\oiii}{}\def\oiii/{[\ion{O\mysp/}{III}]\ 5007\AA}
\newcommand{\vhwhm}{}\def\vhwhm/{V$_{\rm HWHM}$}
\newcommand{\vhwdm}{}\def\vhwdm/{V$_{\rm HW10M}$}

\abstract{We have constructed a grid of photoionization models of spherical, elliptical and bipolar planetary nebulae. Assuming different velocity fields, we have computed line profiles corresponding to different orientations, slit sizes and positions. The atlas is meant both for didactic purposes and for the interpretation of data on real nebulae. As an application, we have shown that line profiles are often degenerate, and that recovering the geometry and velocity field from observations requires lines from ions with different masses and different ionization potentials. We have also shown that the empirical way to measure mass-weighted expansion velocities from observed line widths is reasonably accurate if considering the HWHM. For distant nebulae, entirely covered by the slit, the unknown geometry and orientation do not alter the measured velocities statistically. The atlas is freely accessible from internet. The Cloudy\_3D suite and the associated VISNEB tool are available on request.
}
\addkeyword{methods: numerical}
\addkeyword{line: profiles}
\addkeyword{turbulence}
\addkeyword{planetary nebulae}

\begin{document}

\maketitle

             \section{Introduction}
\label{sec:Intro}
As is known, nebular line profiles bear information on the motions of the emitting gas. The study of line profiles thus allows one to measure the expansion velocity of nebulae and even to determine their internal velocity field. The velocity field is an important key for understanding the dynamics and the genesis of planetary nebulae. The expansion velocity allows one to derive the expansion age, and from this, such parameters as the masses of the central stars \citep{1997A&A...318..256G}. However, the determination of these characteristics is a very difficult task in general, since  only the velocity field perpendicular to the plane of the sky can be observed, through the study of line profiles. 
Even if one could, for a series of ions with different ionization potentials, obtain line profiles in every pixel of a high-resolution image of a PN, one would still need to make assumptions about the projection of the velocity vectors in the plane of the sky, in order to reach a full description of the velocity field. Fortunately, most PNe exhibit some kind of overall regularity in their morphology, and some degree of symmetry (apparently circular, ellipsoidal, bipolar...). So, if an image of the nebula is available, it is possible, with the help of a few assumptions,  to determine the entire nebular geometry and velocity field. Most of the time, however, only limited data are available: line profiles obtained through one slit,  or even, if the object has very small angular dimensions,  line profiles corresponding to the emission of the entire object. If the object is faint, line profiles can be obtained only for the brightest lines, the extreme case being when they are available only for \oiii/. Finally, morphological information on the object may be completely lacking, as is the case for PNe in distant galaxies. In such cases, what is the value of the information that one can derive from line profiles? 

In practice, the analysis of line profiles of planetary nebulae has followed several paths: i) A simple determination of the expansion velocities from observed line widths or from line splitting \citep[most of the references quoted in][]{1989A&AS...78..301W}; ii) Fitting of line profiles with spherical models \citep{GAZ03}; iii) Modeling emission line profiles and Position-Velocity diagrams \citep{1984A&AS...58..273S,1968ApJ...153...49W} and more recently \citet{2005AJ....130.2303M} and \citet{2006RMxAA..42...99S,Stef_Canaries07} which  their powerful tool SHAPE; iv) Fitting of line profiles and other nebular characteristics with a 3D photoionization model \citep{2000ApJ...533..931M,2000ApJ...537..853M,1984A&AS...58..273S}; v) Tomography \citep[][and references therein]{2006A&A...451..937S}.

The question of the robustness of the results is difficult and is almost never addressed in such studies. The atlas of line profiles that we present in this paper is meant as a tool to visualize the variety of possible line profiles, depending on the morphology of the nebula, the velocity field and the observing conditions. The atlas is to be installed in a virtual observatory environment, that will collect in a large data base the models and line profiles computed by any user with the tools we have developed \citep{2006IAUS..234..467M}. 

The description of the atlas is given in Sec.~\ref{sec:Descri}. In the second part of this paper, we use the atlas, in its present stat, for two applications: a search for models reproducing a given profile (Sec.~\ref{sec:appli1}) and an estimation of the robustness of the determination of expansion velocities based on profiles (Sec.~\ref{sec:appli2}).

             \section{Description of the  atlas}
\label{sec:Descri}

We use Cloudy\_3D \citep[][hereafter C3D]{2006IAUS..234..467M}, which is a pseudo-3D photoionization code, similar to the one presented in detail in \citet{MSP05a}, but based on the 1D photoionization code Cloudy \citep{1998PASP..110..761F}  instead of NEBU. The computations for this paper were  based on version 07.02.01 of Cloudy. The size of the cube in which the interpolation of the Cloudy 1D models is done is set to 150$^3$. We run ten Cloudy 1D models in different directions. Once a 3D photoionization model has been obtained, a velocity law is applied to compute line profiles using the section VISNEB of the C3D package, in the same way as described in \citet{2006RMxAA..42..153M}. The profiles are computed on a 55 wavelengths grid covering the range [-60,+60]~\kms/.

The fine structures of the HI, HeI and HeII lines are taken into account by splitting the lines into two components. The relative intensities, velocity shifts and FWHM corrections for the two components are described in Tab.~\ref{tab:finestruc}. The values for \heii/ are adapted from \citet{1999A&AS..135..359C} and P.~Storey (private communication).

\begin{table*}
  \caption{Parameters for the recombination fine structure lines. The global shape is a sum of two gaussian functions. I1 and I2 are the relative intensities of the two components, Shift1 and Shift2 the positions of the center of each gaussian relative to the centroid, Sigma1 and Sigma2 the factor applied to the natural+thermal broadening to take into account the various subcomponents in each gaussian (only for \heii/).}
  \label{tab:finestruc}
  \begin{tabular}{|l|r|r|r|r|r|r|}
    \toprule
    Line & I1 & I2 & Shift1 & Shift2 & Sigma1 & Sigma2 \\
    \midrule
         & relative   & relative   & \kms/  &  \kms/ & \kms/  &  \kms/ \\
    \midrule
    \hbeta/ & 0.41 & 0.59 & 2.7   & -2.0 & 1. & 1. \\
    \hei/   & 0.19 & 0.81 & 13.6  & -3.4 & 1. & 1. \\
    \heii/  & 0.12 & 0.72 & -20.  & 2.9  & 1.08 & 1.20 \\
\bottomrule
  \end{tabular}
\end{table*}

            \subsection{The parameter space and the nomenclature of the models}
\label{sec:ParamSpace}

\begin{figure}
\centering
\includegraphics[width=5.cm]{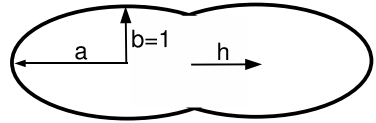}
\caption{Figure defining the geometry and the parameters a and h.\label{fig:morfo}}
\end{figure}

\begin{figure}
\centering
\includegraphics[width=8.cm]{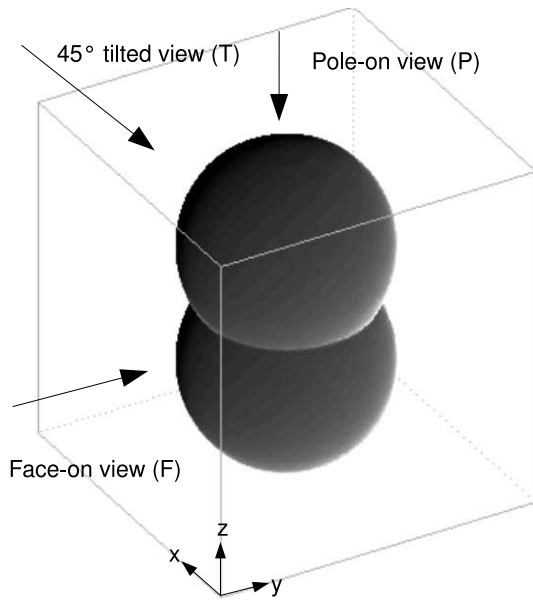}
\caption{Figure defining the x, y, and z axes and showing which is the 
symmetry axis.\label{fig:xyz}}
\end{figure}

All the models in the present version of the atlas have been run by changing the parameters which  have the largest impact on line profiles: i.e. morphology and  velocity law. The models are viewed with different orientations and different slit sizes and locations, all of which strongly affecting the observed line profiles. On the other hand, parameters that play only a secondary or even negligible role have been kept unchanged. This concerns  the effective temperature of the central star (chosen to be T*=10$^5$K, so as to produce low and high ionization ions), the stellar luminosity (set to 10$^4$ L$\odot$), the chemical composition (chosen to be the Cloudy default), the absolute value of the density. In this version of the atlas all the computed models are ionization-bounded.

At each point of the nebula with $r>r_{in}$, the hydrogen density is determined by the following law:

\begin{equation}
  \begin{array}{rcl}
  n_H(r,\theta) & = &  n_H^0 \times  \\ 
  & & \left[ \frac{r_{in}(\theta)}{r_{in}(\theta=0)}  \right]^{-p} \times \\
  & & \exp \left[ -(\frac{r-r_{in}(\theta)}{\Delta r (\theta)})^2  \right] \\
\end{array}
\label{eq:1}
\end{equation}
where $\Delta r(\theta) = 1.6 \times R_{Str}(\theta)$ with $R_{Str}(\theta)$ the Str\"omgren radius in the direction defined by the polar angle $\theta$ ($\theta=0$ in the equatorial direction). The first term of Eq.~(\ref{eq:1}) is the inner equatorial density $n_H^0$ fixed to 2000 cm$^{-3}$, the second term refers to the angular dependency of the inner density using a parameter $p$, and the third term is the gaussian dependency of the radial density. The inner shape of the nebula defined by $r_{in}(\theta)$ is determined by two parameters $a$ and $h$ shown in Fig.~\ref{fig:morfo}, where $a$ is the eccentricity of the ellipsoids and $h$ is the distance between the center of the nebula and the centers of the two lobes (a value of 0.0 for $h$ means a purely ellipsoidal or spherical nebula). 

The adopted nomenclature of the models is the following: 

1rst three digits - generic name of the atlas, in the present version : PN3.

2nd two digits - overall geometry of the inner cavity (defining $r_{in}(\theta)$):  SP: sphere ($a$=1, $h$=0); EL: ellipse ($a$=1.5, $h$=0); BS: bipolar from two spheres ($a$=1, $h$=0.5); BE: bipolar from two ellipses ($a$=1.5, $h$=0.5); BL: bipolar from two elongated ellipses (a=2.2, $h$=0.5).

3rd digit -  parameter $p$ in Eq.~(\ref{eq:1}).

4th digit -  size of the central cavity.  S: small (i.e. corresponding to f=20\% of the Str\"omgren radius in each direction); L: large (i.e. f=80\%).

5th digit -  velocity law. B: ballistic flow (i.e. $v \propto r$); C: constant velocity.

6h digit -  turbulence. 1: turbulence of 10 \kms/; 0: no turbulence.

7th digit -  orientation.  F: face on; P: pole on; T: tilted by 45 degrees (See Fig.~\ref{fig:xyz}).

For example, the model PN3\_BE\_1\_L\_B\_0\_T corresponds to a bipolar nebula obtained from 2 ellipses, with a decrease of the density following the inverse of the inner radius of the nebula, with a large central cavity, expanding with a ballistic velocity law without turbulence, seen tilted by 45 degrees.

In its present version, the atlas is thus composed of 26 photoionization models corresponding to 5 geometries, 3 angular density laws and 2 cavity sizes (the angular density law does not change the spherical models). For each model, four velocity fields are considered. So, in total, there are 104 virtual nebulae, each of which can be observed from 3 different directions (which are identical in the case of spheres, leading to 308 pages for the atlas). 

\begin{figure}
\centering
\includegraphics[width=9.cm]{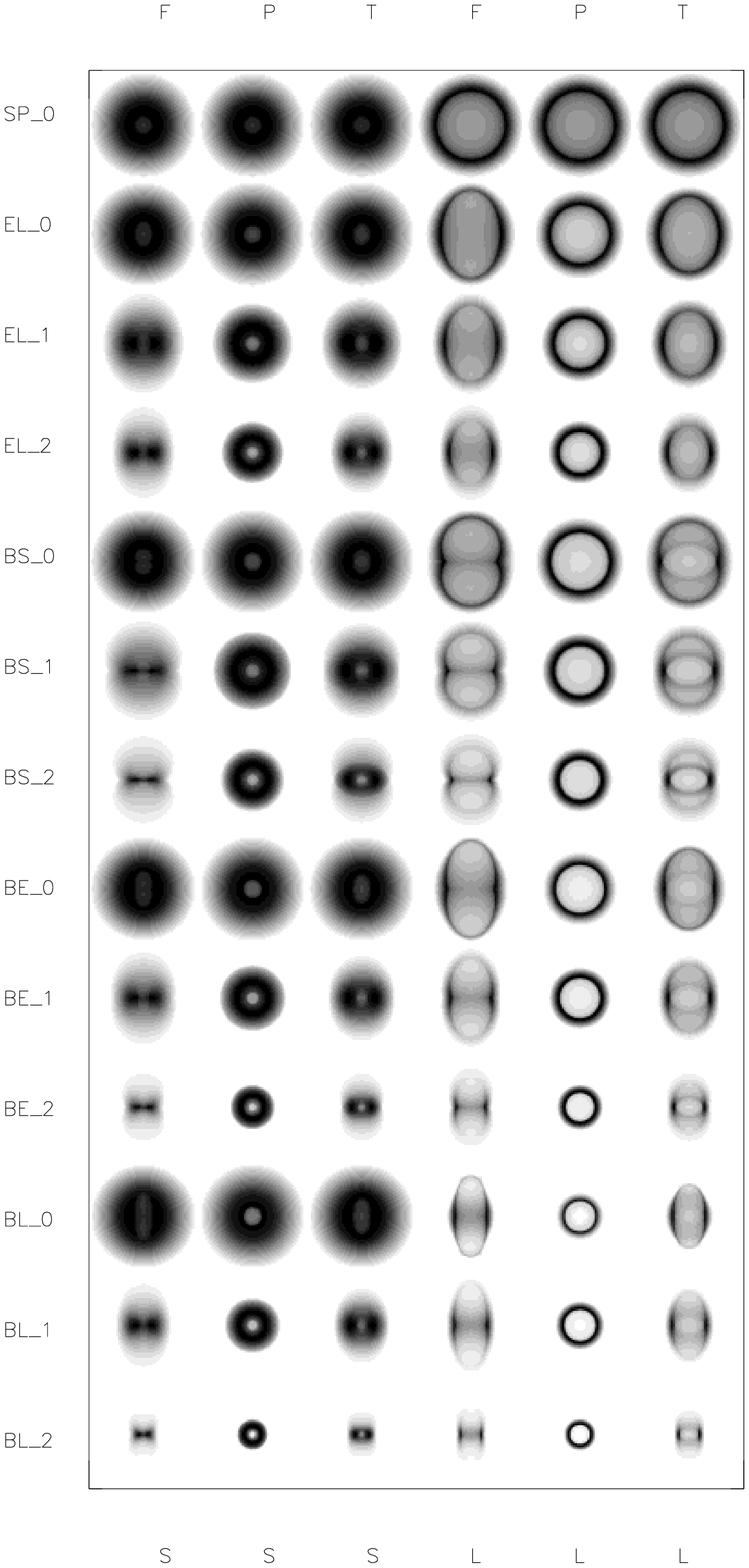}
\caption{The model chart: images of all the models in a the sum of the light in the \hbeta/, \nii/ and \oiii/ lines.\label{fig:map}}
\end{figure}

Figure \ref{fig:map} shows the images of all the models in the atlas, with their various orientations, co-adding the intensities of the \hbeta/, \nii/ and \oiii/ lines. This ''model chart'' allows a quick orientation in the atlas. 

The virtual observations of these models are made through 4 different kinds of apertures: a small square aperture, a long vertical slit, a long horizontal slit, and an aperture covering the entire object. We also compute the case where the aperture is not centered on the object, leading to a total of 7 apertures (the big aperture is not affected by the off-center shift). In all cases, the line profiles were computed with a seeing of 1.5\arcsec. The seeing needs to be considered during the computation of the intensities passing through the selected apertures. On the other hand, we chose to not apply any instrumental profile, as this effect can be taken into account by any user downloading the data from the server.

All the nebulae are set to be at 3~kpc from the observer. On the other hand, the geometrical size of the cube in which the nebula is constructed by C3D is given by the maximum extension of the nebula, which is controlled by the outer radius in the polar direction. This value is changing from one model to the other. In the case of elongated nebulae viewed pole-on, we  basically see the equatorial torus, which appears to be smaller than in the case of spherical nebulae (see Fig.~\ref{fig:map}). Taking the same distance for all the nebulae leads to changes in the ratio pixel/angular size from one model to the other in such a way that the relative size of the slits compared to the size of the equatorial torus is kept constant.
Each virtual observation delivers profiles in 5 lines (namely: \hbeta/, \hei/, \heii/, \nii/ and \oiii/). A total of 10780 different profiles are thus available (resulting from the computation of about two billions individual spectra!).

\subsection{Description of one page of the atlas}

\begin{figure*}
\centering
\includegraphics[width=17.cm]{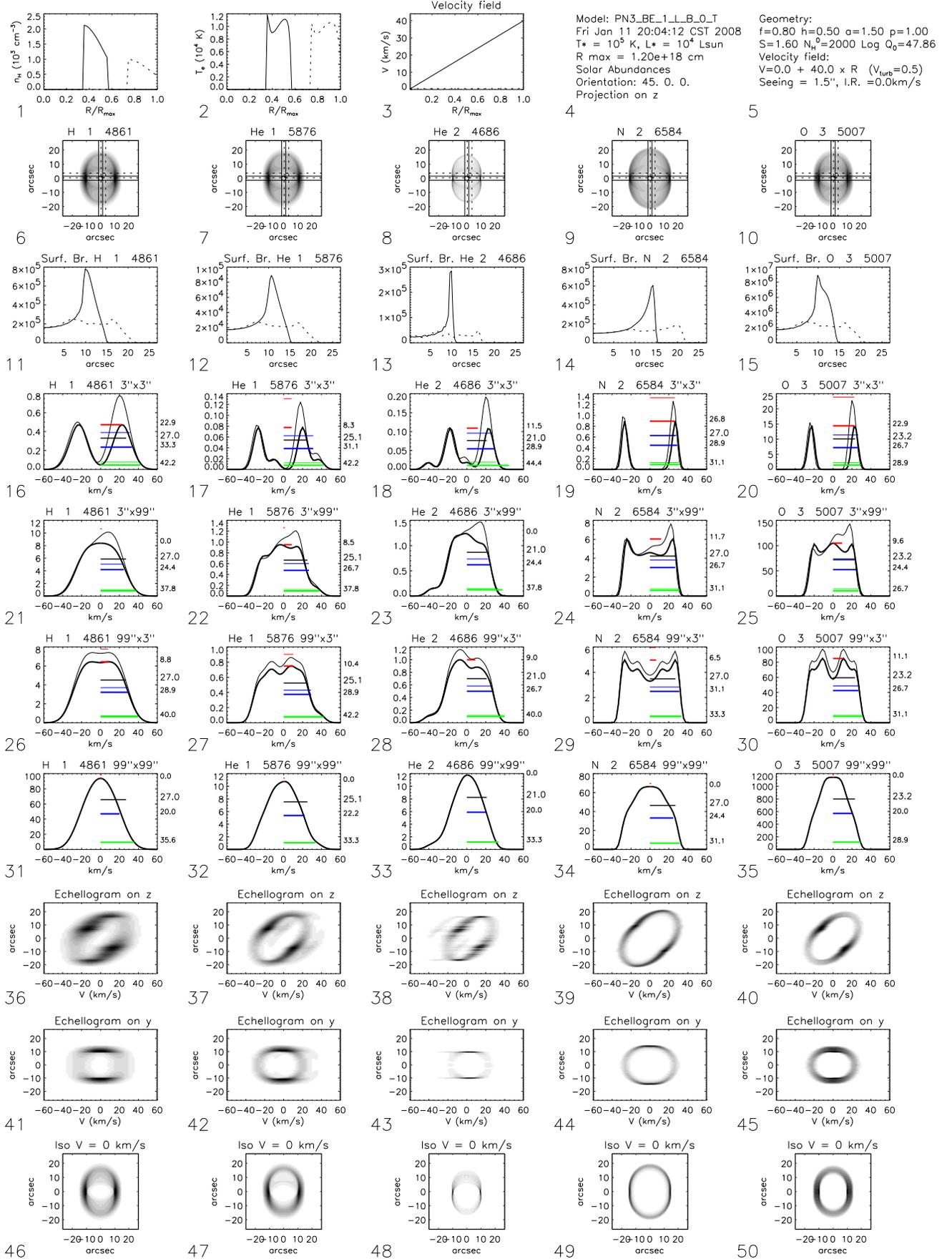}
\caption{One page of the atlas, see text for detailed description.\label{fig:onepage}}
\end{figure*}

Each page of the atlas (an example is given in Fig.~\ref{fig:onepage}) corresponds to a different geometry, velocity law and orientation.  The first row describes the model parameters.  Panel~1 shows the distribution of $n_H$ as a function of $r/r_{max}$ ($r_{max}$ being the half size of the cube, corresponding to the polar extension) along the small axis (continuous line) and along the large axis (dashed line). Panel~2 shows, with the same conventions, the distribution of the electron temperature $T_e$ as computed by C3D. Panel~3 illustrates the assumed velocity field, again with the same conventions. Panels 4 and 5 give additional information on the model definition.

The panels in the second row represent monochromatic images (for the orientation defined in Panel~4) in the lines \hbeta/, \hei/, \heii/, \nii/ and \oiii/. The positions of the slits are also indicated, in solid lines for centrally positioned slits, in dashed lines for non central slits.

In all the following rows, the emission lines considered are the same as in the second row, and they are ordered in the same way, from left to right.

The panels in the third row show the surface brightness distribution in those lines. The continuous curve corresponds to the small axis, the dotted curve to the large axis.

The next four rows show the line profiles, with the bold curve corresponding to a centrally positioned aperture, the thin one corresponding to a non central aperture. The row number 4 (panels 16 through 20) corresponds to a small slit of 3\arcsec$\times3$\arcsec. The row number 5 (panels 21 through 25) corresponds to a horizontal slit of 3\arcsec\ width, while the row number 6 (panels 26 through 30) corresponds to a vertical slit of same width. Finally, the row number 7 (panels 31 through 35) corresponds to an aperture covering the entire nebula. As expected, the line profiles show more complex structures if the slit is small.

Rows 8 and 9 show position velocity diagrams (or echellograms) obtained through a vertical slit centered on the nebula (panels 36 through 40) and through an horizontal slit centered on the nebula (panels 41 through 45).

Finally, row 10 (panels 46 through 50) shows the channel maps of the gas with null radial velocity.

In rows 4 through 7, the horizontal segments indicate the values of the expansion velocities measured by classical methods (for centered slits only). Red: half peak-to-peak velocity. Blue: expansion velocity measured from the half width at half maximum (HWHM). Green: expansion velocity  from the half width at one tenth maximum (HW10M), as given by \citet{1985ApJ...296..390D}. The black segment gives the mass-weighted expansion velocity perpendicular to the plane of the sky (actually the quadratic sum of the expansion velocity, the thermal velocity and the turbulent velocity, weighted by n$_{H^+}$). For each segment the corresponding velocity is also indicated on the right side of the plot, with the mass-weighted velocity using bigger character sizes.

 \subsection{Browsing the atlas}
\label{sec:Browsing}

The entire atlas (images, figures similar to Figs.~\ref{fig:onepage} and \ref{fig:familles} and emission line profiles in ascii format) is available at \url{http://132.248.1.102/Atlas_profiles/}. 

The atlas exhibits a large variety of line profiles. In general, the profiles showing the most complex structures are those obtained through long slits. This is because long slits probe kinetically different regions, while the smoothing effect is minimized. Non central slits always produce asymmetric line profiles. 
In the case of the \hei/ and \heii/ lines, the presence of multiple fine structure components also induces asymmetric profiles, even for symmetric nebula and centered slits.
The most complex line profiles generally correspond to pole-on nebulae. Some of those profiles are reminiscent of those observed by \citet{GAZ03}. As expected, double peak profiles principally occur if observations are made with small slits while large slits give single peak profiles.
In general, ballistic expansion tends to produce triangular profiles (except for \nii/ which is emitted in the outskirts of the nebula and may produce a square profile) while uniform expansion associated with turbulence tends to produce bell-shaped profiles.

However, similar profiles may correspond to very different models, as will be better shown in the next section.

             \section{Which models can account for a given line profile?}
\label{sec:appli1}

So far, in the literature, most of the analysis of line profiles has been either purely empirical \citep[e.g. ][]{2006RMxAA..42...53M} or by fitting to a simple, spherical model \citep[][and references therein]{GAZ03}. However, a large fraction of PNe for which images are available are not spherical. 

One can then wonder how well the velocity field inferred from a model fitting with a spherical nebula represents the true velocity field and whether other solutions, implying non spherical nebulae, may fit the observed data as well. For example, \citet{GAZ03} have invoked the presence of a turbulent velocity field in order to fit their spherical models to the data. 

Our atlas allows us to do the following experiment. We chose one line profile from the atlas, for example obtained through a slit covering the entire object and we search which virtual observations in the atlas give similar line profiles. For this, we define the difference between two profiles $P_1(\lambda)$ and $P_2(\lambda)$ as D$_{1,2} = \sum W(\lambda)\times (P_1(\lambda) - P_2(\lambda))^2 / \sum W(\lambda) $, for example giving the same weight $W(\lambda) = 1$ to all the wavelengths (as apparently done by Gesicki).

\begin{figure}
\centering
\includegraphics[width=8.cm]{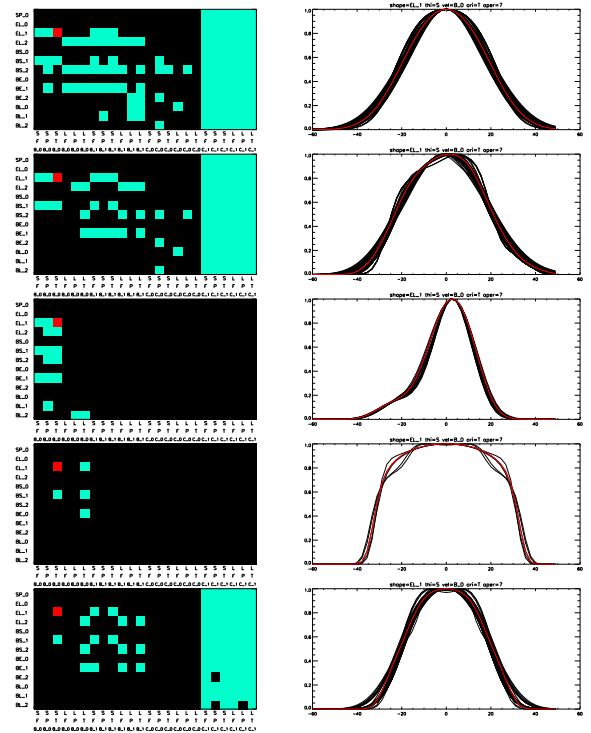}
\caption{Left: In orange: position, in the ``localization map'', of the ``original''  model; in cyan: position, in the ``localization map'' of the models with line profile differing from the original one by by D$_{1,2}<0.001$. Right: the corresponding line profiles. Each row corresponds to a different line, from top to bottom: \hbeta/, \hei/, \heii/, \nii/ and \oiii/.
\label{fig:familles}}
\end{figure}

The result of such an experiment is shown in Fig.~\ref{fig:familles}. Each row corresponds to a different emission line (see caption). The right panels show the profiles (in the same line) that differ by D$_{1,2}<0.001$ from the chosen profile. The left panels indicate to which models these profiles correspond. This is done by assigning to each model the same relative position as in the ''model chart'' shown in Fig.~\ref{fig:map}. The ''model chart'' is repeated four times: from left to right the models B\_0, B\_1, C\_0, and C\_1 are represented (B: ballistic expansion, C: constant velocity, 0: no turbulence and 1: with turbulence ). The red square corresponds to the position of the chosen model, and the green squares correspond to all the models that give a profile that differs only by D$_{1,2}<0.001$ from the chosen one. The result is spectacular. There are, in general, many models that have almost indistinguishable profiles, and these models may have turbulent or non-turbulent velocity fields. If relying on just one line profile, the number of solutions may be very large. If, however, several lines are analyzed at the same time, the number of solutions is considerably reduced. If morphology and line profiles can be analyzed together, the number of solutions is further restricted. 

The main virtue of this exercise is to visualize that the possible solutions may actually be quite numerous. Because of the very coarse grid of models in the atlas, we obviously miss many other combinations of parameters. This shows that any solution obtained by model fitting cannot be regarded as necessarily correct. We emphasize that all of our models have very simple density laws and geometries. If we were to consider more complex morphologies the number of solutions would grow accordingly. 

Notice that no instrumental profile has been applied, so that  the user can applyhis own instrumental profile to themodeldata available from the server. Taking into account the instrumental profile would  increase the number of profiles similar to a given one.

It is straightforward, with C3D, to perform the same experiment for different velocity laws, and thus to visualize the different solutions corresponding to an observed line profile. 
             \section{How good are simple recipes to estimate expansion velocities?}
\label{sec:appli2}

In many instances, either because there is no spatial information on the nebula or because it would be too time consuming to make an appropriate line profile fitting for a sample of nebulae taking into account all possible solutions, one relies on simple formulas to estimate the expansion velocity.

Taking our atlas as a reasonable representation of the different possible cases in nature, we can estimate the errors and biases in the  formulas commonly used to estimate expansion velocities. The distribution of morphologies of real PNs is obviously not comparable to the one we adopt here, but the global results of this work are still valid.

\begin{figure*}
\centering
\includegraphics[width=14.cm]{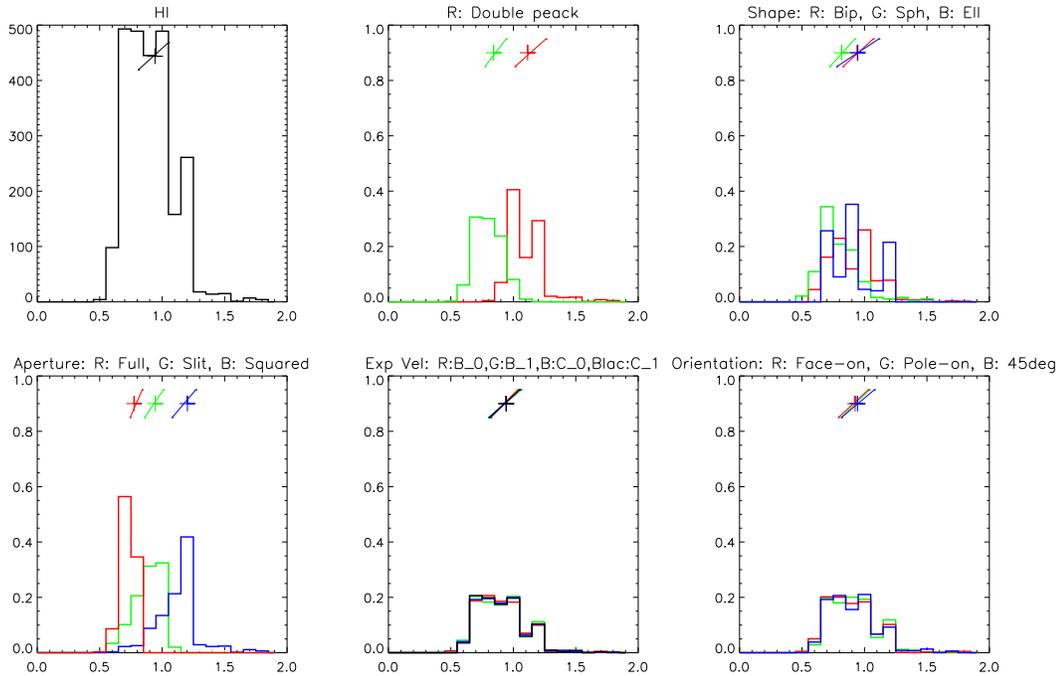}
\caption{Histograms for the HWHM-velocity determination normalized to the mass-weighted real values, for the \hbeta/ line. See text for a detailed description\label{fig:histo1}}
\end{figure*}

To do this, we construct histograms of the HWHM velocities (expressed in units of mass-weighted mean velocities, $v_M$), grouping the models according to  various criteria.  Figure~\ref{fig:histo1} upper left  shows the histogram for all the virtual observations of the \hbeta/ line.  In Fig.~\ref{fig:histo1} upper middle, the virtual observations are grouped into double peak profiles (red) and single peak ones (green). In Fig.~\ref{fig:histo1} upper right, the models are grouped by morphology: spherical ones are in green, bipolar ones are in red, elliptical ones are in black. In Fig.~\ref{fig:histo1} lower left, the grouping is by apertures: red is for apertures covering the entire objects, green is for long slit apertures, blue is for small apertures, whether the aperture is centered or off-center. Fig.~\ref{fig:histo1} lower middle distinguishes objects with a various velocity fields. Finally, Fig.~\ref{fig:histo1} lower right distinguishes the nebulae by orientation: face-on ones are in red, pole-on are in green, tilted are in blue. In the top of each panel, the crosses indicate the median value, and the extremities of the inclined segment indicate the quartiles, the color of each cross being the same as that of the corresponding  histogram.

It is striking that, in all the cases, the median values are displaced from 1 by less than 20\%  and the quartiles extend to 10 -- 15\% from the median. This indicates that such a simple estimate as HWHM is actually a quite accurate estimate of the mass-weighted expansion velocity.

\begin{table*}
  \caption{Velocity correction coefficients.}
  \label{tab:correc}
  \begin{tabular}{|l|r|r|r|r|r|r|}
    \toprule
    Line & All & Single Peak & Double Peak & Full object & Long Slit & Small Aperture \\
\midrule
\multicolumn{7}{|c|}{HWHM}\\
\midrule
   \hbeta/&  0.94 $_{ -0.13}^{+  0.11}$&   0.84 $_{ -0.07}^{+  0.10}$&   1.12 $_{ -0.10}^{+  0.15}$&   0.77 $_{ -0.03}^{+  0.06}$&   0.94 $_{ -0.09}^{+  0.06}$&
   1.20 $_{ -0.12}^{+  0.07}$\\ 
\midrule
     \hei/&  1.05 $_{ -0.14}^{+  0.10}$&   1.02 $_{ -0.16}^{+  0.09}$&   1.15 $_{ -0.13}^{+  0.18}$&   0.86 $_{ -0.04}^{+  0.16}$&   1.05 $_{ -0.09}^{+  0.07}$&
   1.30 $_{ -0.14}^{+  0.04}$\\ 
\midrule
    \heii/&  1.10 $_{ -0.09}^{+  0.07}$&   1.02 $_{ -0.07}^{+  0.09}$&   1.21 $_{ -0.09}^{+  0.13}$&   0.96 $_{ -0.02}^{+  0.05}$&   1.10 $_{ -0.07}^{+  0.04}$&
   1.30 $_{ -0.14}^{+  0.13}$\\ 
\midrule
     \nii/&  1.05 $_{ -0.11}^{+  0.03}$&   0.88 $_{ -0.10}^{+  0.07}$&   1.05 $_{ -0.05}^{+  0.04}$&   0.88 $_{ -0.06}^{+  0.07}$&   1.05 $_{ -0.07}^{+  0.03}$&
   1.12 $_{ -0.06}^{+  0.06}$\\ 
\midrule
    \oiii/&  0.98 $_{ -0.09}^{+  0.08}$&   0.85 $_{ -0.09}^{+  0.10}$&   1.05 $_{ -0.09}^{+  0.06}$&   0.86 $_{ -0.07}^{+  0.09}$&   0.98 $_{ -0.05}^{+  0.07}$&
   1.16 $_{ -0.10}^{+  0.02}$\\ 
\midrule
\multicolumn{7}{|c|}{HW10M}\\
\midrule
   \hbeta/&  1.43 $_{ -0.08}^{+  0.07}$&   1.37 $_{ -0.09}^{+  0.08}$&   1.51 $_{ -0.08}^{+  0.12}$&   1.28 $_{ -0.02}^{+  0.09}$&   1.43 $_{ -0.08}^{+  0.03}$&
   1.60 $_{ -0.09}^{+  0.12}$\\ 
\midrule
     \hei/&  1.56 $_{ -0.12}^{+  0.06}$&   1.53 $_{ -0.10}^{+  0.09}$&   1.58 $_{ -0.13}^{+  0.23}$&   1.43 $_{ -0.09}^{+  0.06}$&   1.58 $_{ -0.11}^{+  0.04}$&
   1.75 $_{ -0.15}^{+  0.07}$\\ 
\midrule
    \heii/&  1.81 $_{ -0.11}^{+  0.18}$&   1.79 $_{ -0.17}^{+  0.19}$&   1.84 $_{ -0.09}^{+  0.14}$&   1.61 $_{ -0.03}^{+  0.21}$&   1.79 $_{ -0.08}^{+  0.13}$&
   1.98 $_{ -0.08}^{+  0.14}$\\ 
\midrule
     \nii/&  1.23 $_{ -0.09}^{+  0.14}$&   1.28 $_{ -0.13}^{+  0.07}$&   1.18 $_{ -0.03}^{+  0.20}$&   1.17 $_{ -0.12}^{+  0.11}$&   1.27 $_{ -0.13}^{+  0.10}$&
   1.31 $_{ -0.17}^{+  0.26}$\\ 
\midrule
    \oiii/&  1.28 $_{ -0.13}^{+  0.10}$&   1.28 $_{ -0.12}^{+  0.09}$&   1.28 $_{ -0.13}^{+  0.12}$&   1.21 $_{ -0.16}^{+  0.07}$&   1.30 $_{ -0.15}^{+  0.08}$&
   1.40 $_{ -0.26}^{+  0.17}$\\ 
\bottomrule
  \end{tabular}
\end{table*}

\begin{figure*}
\centering
\includegraphics[width=14.cm]{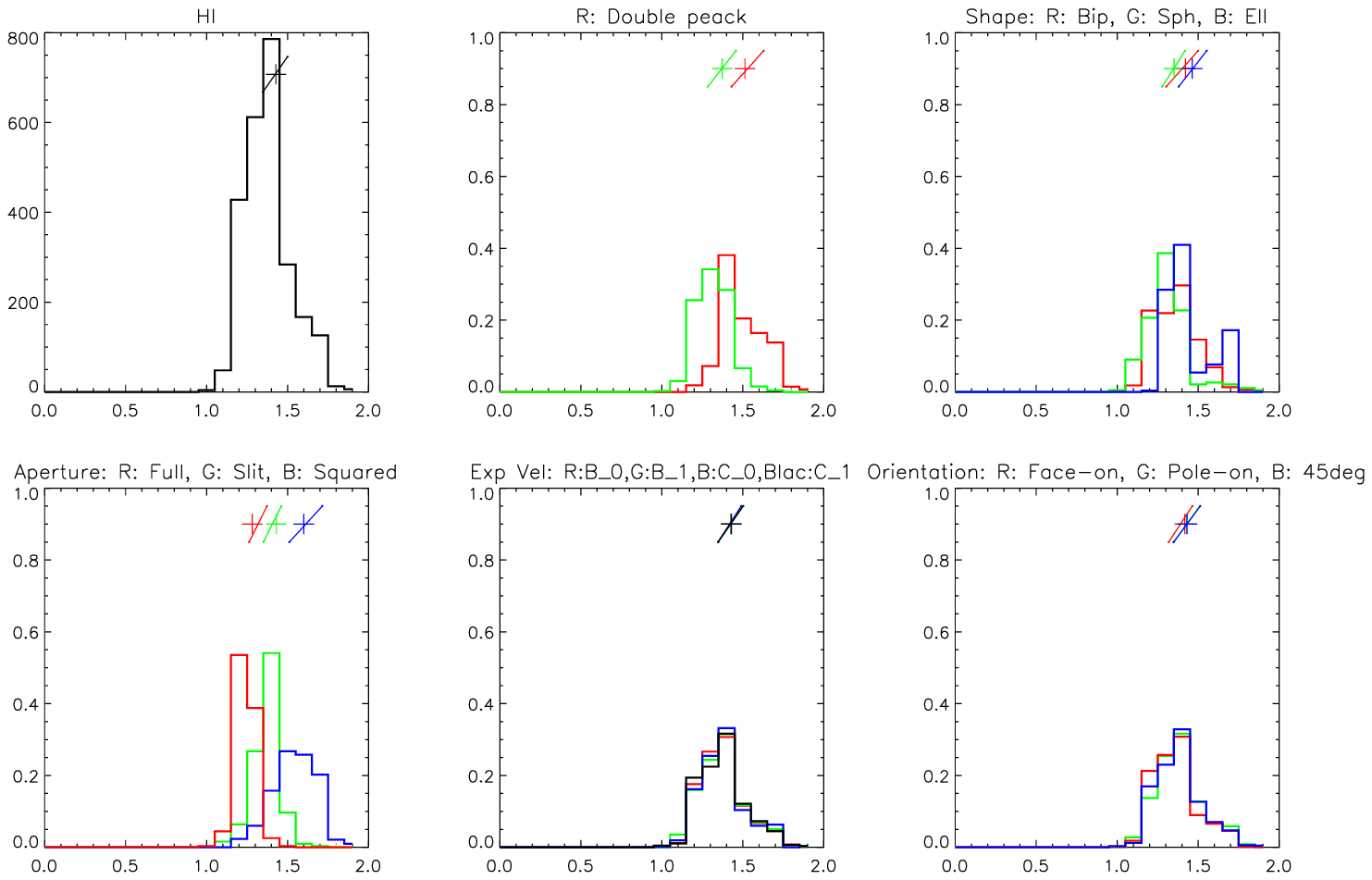}
\caption{As Fig.~\ref{fig:histo1} but for HW10M-velocities.\label{fig:histo2}}
\end{figure*}

Figure~\ref{fig:histo2} is analogous to Fig.~\ref{fig:histo1}, but now with half-width at 1/10 maximum (HW10M) velocities instead of HWHM. 
While the histograms are again rather peaked, we note that the median value is significantly larger than 1 (typically 1.2 -- 1.3). Therefore, the HW10M velocity gives a more biased estimate of  $v_M$. Since \vhwdm/ is further plagued by the influence of noise to determine the position of the continuum, we conclude that \vhwhm/ is to be preferred  to estimate the expansion velocity field. 

The parameters that have a large influence on the velocity determination are the ones that produce different mean values. We find that the global shape of the profile (single or double peak) and the relative size of the aperture (small square, long slit, whole object) clearly affectes the inferred expansion velocity. On the other hand, for HWHM  velocities, neither the real morphology of the nebula, nor the expansion velocity law, nor the orientation of the nebula have a strong influence on the velocity determination. The latter parameters are those which are not known from the observer, while the first ones are always determined. These results attest the reliability of simple velocity estimator HWHM in the sense that invisible parameters do not bias the determined values. 

Table~\ref{tab:correc} lists the ratios of the velocities determined by the HWHM and HW10M methods relative to the ``true'' values (mass-weighted velocities), in various cases depending on the shape of the profile (single or double peaked) or the relative size of the aperture used. The ``errors'' correspond to the 25 and 75 percentiles. It is important to notice that this list of ratios is based on histograms constructed using the models from our atlas, where no attempt is made to reproduce the real distribution of PN morphologies and that those numbers are only indicative of the effects of various parameters but are not to be taken blindly as correction coefficients to apply to real observations.

The effect of line splitting due to fine structure is obvious in the case of \hei/ and \heii/ lines, where the blueshifted low intensity component of the multiplet is still intense enough to be taken into account by the HW10M method. The determined expansion velocities are then determined 80\% higher than the true values, while for unsplitted lines they are overestimated by 20 to 30\%.  

The general conclusion  is that expansion velocities are reasonably well estimated from \vhwhm/, especially if one applies the correction given in Tab.~\ref{tab:correc}.

             \section{Summary}
\label{sec:conclusion}

Using the pseudo-3D photoionization code Cloudy\_3D, we have constructed a grid of photoionization models of planetary nebulae with different morphologies, including elliptical and bipolar geometries. We have then assumed different expansion laws, and computed line profiles corresponding to different orientation of the models considered as virtual nebulae, and to different slit sizes and positions. The resulting atlas is available at \url{http://132.248.1.102/Atlas_profiles/}. In its present version, it is already useful to visualize the variety of profiles that can be obtained according to the physical properties of the nebulae and to the observing conditions, and to help in the interpretation of observational data.

We have shown that in many circumstances, line profiles are degenerate, and the recovery of the true geometry and velocity field from observations requires lines from ions with different masses and different ionization potentials.

We have also shown that the empirical way to measure mass-weighted expansion velocities from the HWHM of observed line profiles is accurate within about 20\% in the  range of geometries and velocity fields considered in the present version of the atlas. In the case of observations of distant nebulae which are entirely covered by the slit, the unknown geometry and orientation of the nebula do not alter the measured velocities, when considered statistically.

We plan to integrate our atlas in a virtual observatory environment. For the time being, the Cloudy\_3D suite and the associated VISNEB tool are available on request from C.M. and may be used as a help to interpret observed expansion profiles.

One application of the atlas will be to compare the synthetic line profiles and PV-diagrams with observed data such as the SPM Kinematic Catalogue of Planetary Nebulae \citep{2006RMxAC..26..161L}.

\acknowledgements

We are grateful to Gary Ferland for having made public the source of the code Cloudy on which C3D is based. We thanks our referee, Franco Sabbadin, for detecting a numerical error in a previous version of the atlas and for pointing out the effect of fine structure line splitting.
The computations were carried out on a AMD-64bit computer financed by grant PAPIIT IX125304 from DGAPA (UNAM, Mexico). C.M. is partly supported by grants Conacyt-49737 and PAPIIT-IN115807 (Mexico). G.S. is grateful to the Instituto de Astronomia, UNAM, Mexico, for hospitality and financial support (grant Conacyt-49737 and DGAPA IN-118405).



\end{document}